\documentclass[showpacs,twocolumn,amsmath,amssymb,prb,eqsecnumbers]{revtex4}
\usepackage{epsfig,color}

\usepackage{graphicx}
\usepackage{dcolumn}
\usepackage{bm}
\usepackage{graphicx}

\def\mb{\mathbf}

\newcommand{\eps}{\varepsilon}

\newcommand{\nn}{\nonumber}
\newcommand{\beq}{\begin{equation}}
\newcommand{\be}{\begin{equation}}
\newcommand{\eeq}{\end{equation}}
\newcommand{\ee}{\end{equation}}
\newcommand{\bea}{\begin{eqnarray}}
\newcommand{\eea}{\end{eqnarray}}
\newcommand{\bwt}{\begin{widetext}}
\newcommand{\ewt}{\end{widetext}}

\begin{document}

\title{Single 20\,meV boson mode in KFe$_2$As$_2$ detected
by point-contact spectroscopy}

\author{Yu.G.\ Naidyuk, O.E.\ Kvitnitskaya, N.V.\ Gamayunova}
\affiliation{B. Verkin Institute for Low Temperature Physics and Engineering,
National Academy of Sciences of Ukraine, 47 Lenin Ave., 61103, Kharkiv,
Ukraine}
\author{L.\ Boeri}
\affiliation{ Technische Universit\"at Graz, 5150
Institut f\"ur Theoretische Physik-Computational Physics
Petersgasse 16/II, 8010 Graz,
Austria}
\author{S.\ Aswartham, S.\ Wurmehl, B. B\"uchner, D.V.\ Efremov, G.\
Fuchs, S.-L.\ Drechsler}
\affiliation{Leibniz-Institut f\"ur Festk\"orper-
und Werkstoffforschung Dresden e.V.,Postfach 270116, D-01171 Dresden,
Germany}

\begin{abstract}
We report an experimental and theoretical investigation of the electron-boson
interaction in KFe$_2$As$_2$ by point-contact (PC) spectroscopy,
model, and \textit{ab-initio} LDA-based  calculations for the
standard electron-phonon Eliashberg function.
The PC spectrum viz. the second derivative of the $I-V$ characteristic of
representative PC exhibits a pronounced maximum at about 20\,meV and
surprisingly a featureless behavior at lower and higher energies.
We discuss phonon and non-phonon (excitonic) mechanisms
for the origin of this peak. Analysis of the underlying source of this
peak may be important for the understanding of serious puzzles
of superconductivity in this type of compounds.
\end{abstract}


\date{today}

\maketitle

\section{INTRODUCTION}

The superconductivity in iron-pnictides and chalchogenides (FeSC),
and in particular the doping, pressure, and disorder dependencies
of the superconducting critical temperature $T_c$
are still under lively discussions since their discovery
more than five years ago.
More importantly, the question of the nature of the bosonic glue for Cooper
pair-mediated superconductivity is still open.

In order to estimate the conventional contribution of phonons to the Cooper pairing,
immediately after  the discovery of FeSC, the spectral function of the electron-phonon
interaction (EPI) has been calculated from first principles for a number of FeSC by
different authors \cite{Boeri2008,Subedi2008,Mazin2008,Yildirim2009}.
All calculations showed that the EPI is not strong enough to get  $T_c$ values
exceeding a few K. Later it was shown that  considering a magnetic or paramagnetic
ground state, i.e.\ a state with large local magnetic moments at the Fe sites,
leads to a $\sim$ 50 $\%$ enhancement of the electron-phonon (EP) coupling.
\cite{Boeri2010,Yndurain2011}  However, even this effect is not enough to yield critical
temperatures above a few K, and is thus insufficient to explain the high $T_c$ of FeSC.
Based on these finding, pure phonons were excluded as a leading glue for Cooper pairing.

Currently, two main scenarios  for Cooper pairing in the pnictides are under debate.
The first is the inter-band spin-fluctuation scenario which, at optimal doping,
favors s$_\pm$-wave superconductivity, characterized by gap functions with opposite
signs on the electron ($M$-point centered) and hole ($\Gamma$-point centered)
 surface sheets. \cite{Mazin2008,Kuroki2008}
The second scenario is based on the interplay of orbital fluctuations with conventional
electron-phonon coupling.
%LB: Reformulated the sentence, hopefully without changing the meaning.
In this case superconductivity exhibits ordinary s$_{++}$-wave symmetry, i.e.\
the gap functions show the same sign on the electron and hole Fermi surface sheets \cite{Kontani2012}.
For the transition regime between these limiting cases due to disorder
see Ref.\ \onlinecite{Efremov2011}.
In this scenario, a small bare EP coupling constant ($\lambda_{\rm ph} \approx $ 0.2)
can be so strongly enhanced by orbital fluctuations to  cause $s_{++}$-wave superconductivity
with reasonable critical temperatures.
Given the discrepancy between the various theoretical scenarios, it is highly desirable
to obtain an independent estimate of the actual coupling of electrons to various bosonic
excitations from an experimental source.

Point-contact (PC) spectroscopy (PCS) \cite{Naidyuk2005} is one of the few available tools
to address this question, because it permits to measure the spectral function for the
interaction of conduction electrons with different types of bosonic excitations: phonons,
%ferro- and antiferromagnetic
paramagnons (spin fluctuations), crystal-electric field excitations, etc.
In particular, several authors have recently underlined the potential importance of PCS
in iron pnictides and chalchogenides, as a tool to identify new featuress due to the
interplay of strong electronic correlations with spin and orbital fluctuations,
close to an orbital-selective Mott transition.~\cite{Arham12,Greene12,Gonnelli2013}.

In PCS, the second derivative of the $I-V$ curves of  the ballistic PC, in other words the PC spectrum,
represents directly the spectral function $\alpha^2_{\rm PC}F(\omega)$
of the interaction of conduction electrons with phonons or other bosonic excitations (electron-boson
interaction, EBI).\cite{Kulik1977,Kulik1992,Jansen1980,footnote2}.

In order to obtain reliable results the measurements have to be performed at temperatures
considerably lower than the characteristic energy of the bosonic excitations. The underdoped
compound KFe$_2$As$_2$ (K122), with a low critical temperature $T_c \leq 4$\,K, is thus one of
the best candidates in the 122 family of FeSC for studying of PC spectra in the normal state.
It is also the only member of this family where nodal superconductivity, probably of
$d$-wave nature, has been reported. \cite{Reid2012,Abdel2013,Grinenko2014}

In this paper we present a combined experimental and theoretical study of the PC spectra
of K122. We find that the PC spectrum
%or the second derivative of the $I-V$ curves
as a function of  $\omega = eV$ grows nearly linearly at small $\omega$, has a maximum
at $\sim 20$\,meV, and then decays  as $1/\sqrt{\omega}$.
Such a behavior can hardly be attributed to phonons or pure spin fluctuations. Based on
a simplified analytical model of the underlying electronic structure, we propose that
this feature is due to excitonic charge excitations.

The outline of the paper is the following. After a short introduction into the theory of PCS,
we present our  PC spectra and discuss them first in terms of a standard EP model.
We then show the outcome of a linear response~\cite{dfpt,qe} calculation for the EPI spectrum
of K122, which shows a relatively featureless  spectrum, with a very weak total EPI.
This  is incompatible with the measured PC spectrum.
In the following section we introduce a novel scenario involving charge excitations into the
empty electron pockets. We conclude discussing the physical consequences
of our findings and prospects for further work.

\section{PCS OF BOSONIC EXCITATIONS}

PCS is a direct tool to study the EBI. According to the general theory of PCS, \cite{Kulik1977,Jansen1980}
the second derivative of the $I-V$ curve in case of a {\it ballistic} PC is given by
the energy derivative of the scattering rate.
Among the various contributions to the scattering rate, inelastic spin and charge interband
scattering, as well as diagonal and off-diagonal EBI can be identified.
First, we consider the most common case of EBI. In this case the second derivative of the $I-V$
curve $R^{-1}dR/dV=R^{-2}d^2V/dI^2$ of the ballistic PC is directly proportional to the electron-boson spectral function
$\alpha^2_{\rm PC}F(\omega )$ \cite{Kulik1992}:
\begin{equation}
\frac{1}{R} \frac{dR}{dV} = \frac{8ed}{3\hbar v_F} \alpha^2
_{\rm PC}F(\omega)|_{\omega= eV} ,
\end{equation}
where $R= dV/dI$ is the differential PC resistance, $e$ is the electron charge,
$d$ is the PC diameter and $v_F$ is the Fermi velocity. Hence, we yield:

\begin{equation}
\alpha^2_{\rm PC}F(\omega)= \frac{3}{8} \frac{\hbar v_F}{ed} R^{-2} \frac{d^2 V}{dI^2}\propto \frac{d^2 V}{dI^2}.
\end{equation}

The spectral information can be extracted if the size $d$ of PC (i.e.\
the contact area with the K122 single crystal) is less than the elastic ($l_{el}$)
and inelastic ($l_{in}$) mean free path of electrons ($d \ll l_{el}, l_{in}$),
i.e. in the case of ballistic contacts. Spectroscopy is also possible under a less strict
condition, i.e.\ also in the case that only ($l_{in}$) is larger than the contact size $d$.
This is,  so-called, diffusive regime  ($l_{el} \ll d \ll \sqrt{l_{el} l_{in}}$).
In both cases electrons are accelerated in the PC up to a maximum energy $\omega$=$eV$,
and varying the applied voltage allows energy-resolved spectroscopy. The opposite limit is the
thermal regime ($\min (l_{in},  \sqrt{l_{el} l_{in}} \ll d$) of the current flow. In this case
the electron transport behaves like in the bulk material, which results in Joule heating: the
temperature in the PC core increases with the applied bias voltage \cite{Naidyuk2005,Verkin1993}.

It is important to point out that the PC EBI function $\alpha^2_{\rm PC}F(\omega)$ differs
from the Eliashberg
%LB: is a sort of transport spectral functions, which
thermodynamical EBI function $\alpha^2 F(\omega)$ by a factor due to the kinematic restrictions
for the electron scattering processes in the contact. For spherical Fermi surfaces, the PC spectral
functions are obtained averaging the $\mb{k}$-dependent thermodynamical spectral functions
$\alpha^2 F({\bf k,k'},\omega)$ with a weighting factor 1/2(1-${\theta}$/tan${\theta}$) over
the angle  $\theta$ formed by the incoming ({\bf k}) and outgoing ({\bf k}') momenta of the electrons.
%({\bf please check!!!})%
Evaluating  the corresponding average for K122, which possesses multiple and highly
anisotropic Fermi surface sheets, is a non-trivial task which deserves a theoretical study by itself,
and is therefore beyond the scope of the present work.
%~\cite{Gonnelli2010}

PCS can provide useful information on the boson modes coupled to electrons, even without a detailed
knowledge of the electron scattering processes taking place in the contact \cite{Naidyuk2005,Kulik1977,Kulik1992}.
In fact, this is a very powerful tool to identify candidates for the pairing glue and for the mass
renormalization observed in the electronic specific heat in the normal state at low temperatures.
In principle, it is  possible to extract the EBI functions also from PC spectra in the superconducting state.
However, the features of the PC spectra induced by the superconducting state, such
as the Andreev-reflection or the critical current (self-magnetic field) effects, are much stronger
than those due to EBI, at least in the energy range of the order of superconducting gap value.
Therefore, the EBI spectral function is usually extracted from PC spectra measured in the normal state,
that is, above $T_c$ or above the upper critical field $H_{c2}$ in the case of type-II superconductors.
%LB: which we considered here.

In FeSC, PC  measurements above $T_c$ (typically in the range of 30--55\,K) will yield little information,
since the energy resolution in the PCS \cite{Naidyuk2005} decreases with increasing temperature and amounts to
about 20\,meV at 40\,K. To circumvent this difficulty, we decided to carry out our first measurements on
%first on a FeSC with a low $T_c$ like
K122, where $T_c \lesssim 4$\,K. We expect that, although $T_c$ in this compound is low in comparison
with $T_c \sim$\,40\,K for optimally doped 122 FeSC,
%the PCS data on KFe$_2$As$_2$ could help us to identify the
the main bosonic features in the PC spectra
%and that these could be
could be similar in systems with a lower hole doping, closer to the magnetic parent compounds
(Ba-112 or Sr-122). This is partly confirmed by our first-principles calculations, which show
a quite similar distribution of the EPI  for K122 and Ba122 (compare Fig.\,\ref{fig4} and Ref.~\onlinecite{Boeri2010}).
A further advantage of K122 is the large electron mean free path, compared to that of doped FeSC,
due to a lower amount of disorder, that could makes it easier to realize ballistic contacts required by the PCS.\cite{Naidyuk2005,Kulik1977,Kulik1992}.

\section{EXPERIMENTAL DETAILS}

\begin{figure}[b]
\includegraphics[width=0.45\textwidth]{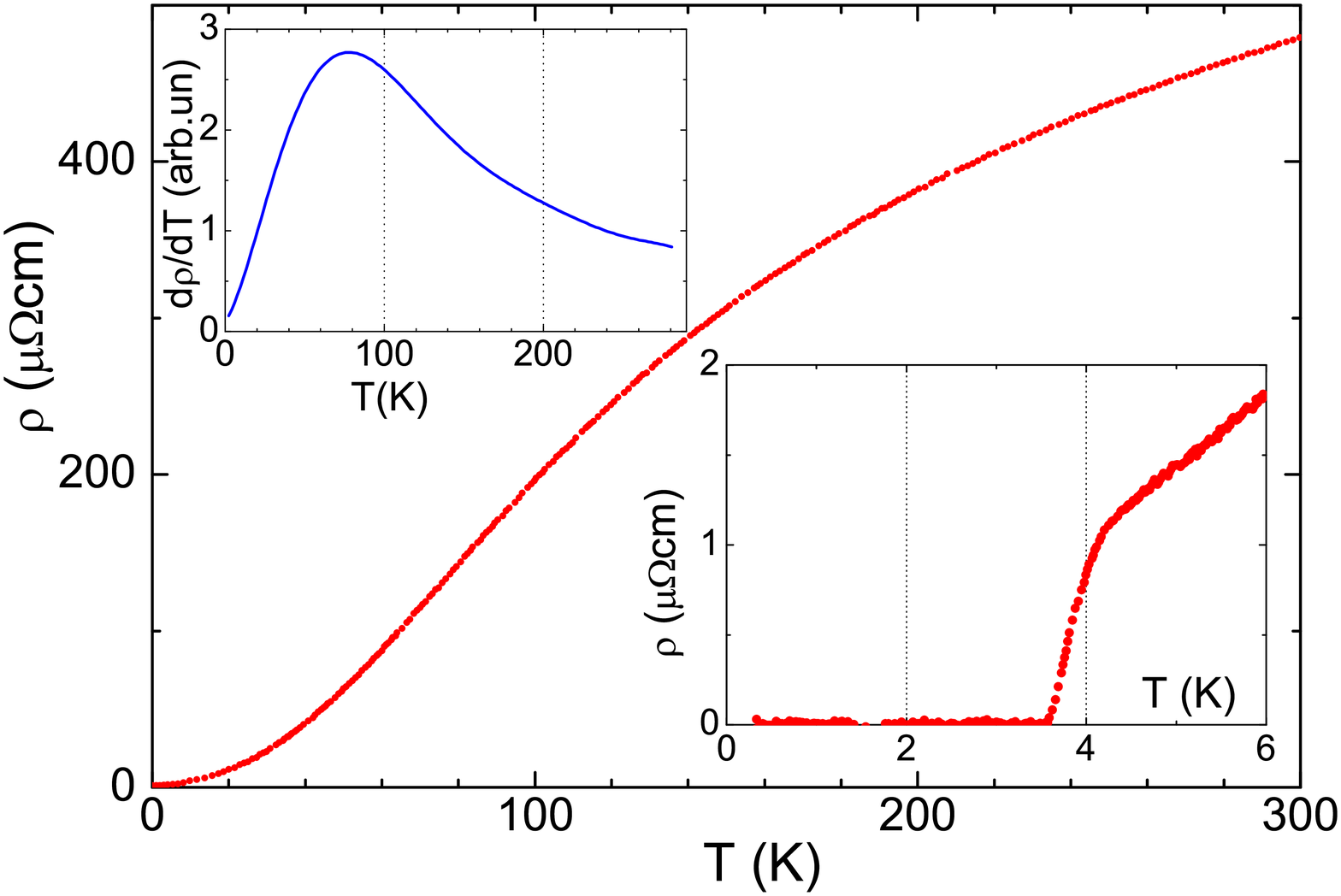}
\caption{(Color online) Resistivity of one  of the K122 single crystals used in our experiments.
The residual resistivity ratio of all crystals amounts to about 400.
Left inset: shape of the temperature derivative of the resistivity in the main panel.
Right inset: the superconducting transition as seen in the resistivity data.
\label{fig:fig1}}
\end{figure}

High-quality K122 single crystals were grown using the self-flux method as described in Refs.\, \cite{Abdel2012,Abdel2013}.
%{\bf is there a part missing in the previous sentence, or is only one sample/batch of samples grown with KAs flux?}
Their lateral dimension was as large as $1\times0.5$ mm$^2$ and the thickness is up to 0.1 mm.
The onset of the superconducting  transition is slightly below 4\,K (see Fig.\,\ref{fig:fig1}, right inset).
The temperature dependence of the resistivity has a typical metallic behavior with low residual
resistivity (see Fig.\,\ref{fig:fig1}) and quantitatively it is similar to that of Ref.~\onlinecite{Rotter2008},
where the high-$T_c$ superconductivity in 122 FeSC was reported for the first time. The
PCs were established by the conventional {\textquotedblleft}needle-anvil{\textquotedblright}
technique \cite{Naidyuk2005}, touching a cleaved single crystal surface with a sharpened thin
Cu or Ag wire. A first set of measurements with a Cu needle has been carried out at the IFW (Dresden),
and a second set of measurements with an Ag needle have been performed on a different batch
of single K122 crystals in the ILTPE (Kharkiv). The differential resistance signal $V_1\propto dV/dI(V)= R(V)$
and the second derivative signal $V_2 \propto d^2V/dI^2(V)$, were
recorded by sweeping the dc current $I$ on which a small ac current $i$ was superimposed
using the standard lock-in technique. Here $V_1$ and $V_2$ are the $rms$ amplitudes of the first
and the second harmonics of the modulating signal, respectively, such that $V_1$ and $V_2$
are related by $V_2 = 8^{-1/2} V_1(dV_1/dV)$.
The measurements of our PC spectra were performed in most cases at 4\,K.
%\textcolor{blue}{\bf Yura, please check the temperature! We have replaced "3" K by "4" K in order
% to stay definitly on the normal state side as discussed above.}
%in some cases the temperature was varied in the range in between 2--10~K
%to search for superconducting features in the spectra.

\section{RESULTS AND DISCUSSIONS}
\subsection{RESULTS}
We have measured the $V_2 \propto d^2V/dI^2$ characteristics for more than one hundred K122--Cu
(or Ag) PCs. We can distinguish mainly two types of PC spectra.
\begin{figure}[t]
\includegraphics[width=0.45\textwidth]{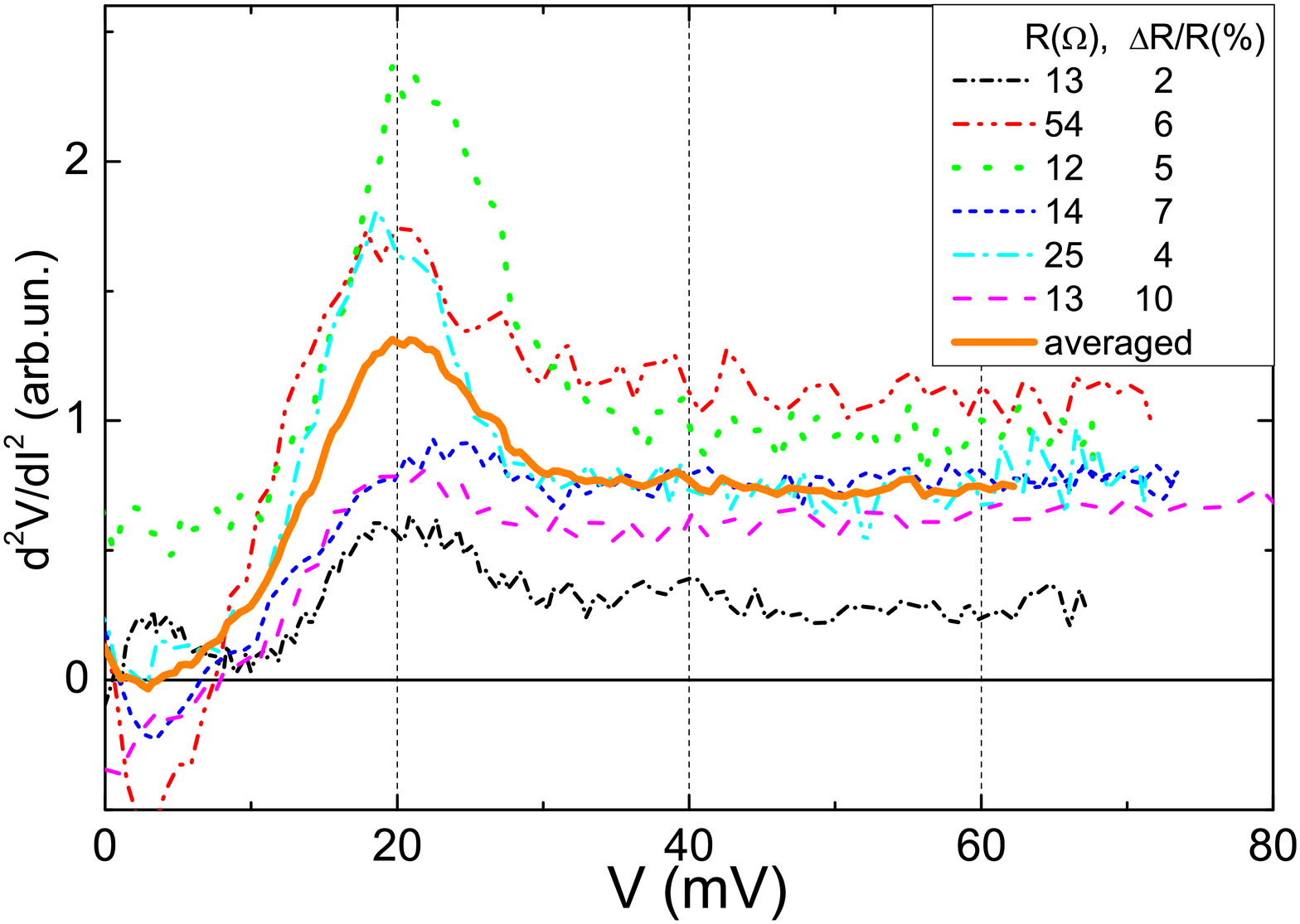}
\caption{
(Color online) PC spectra measured for 6 selected K122--Cu PCs with different resistance. The thick (orange) curve is
the average of six presented  spectra. The inset shows the PC resistance for each curve in the main panel
and the change of the differential resistance by a voltage increase up to the end of the spectrum.
\label{fig:fig2}}
\end{figure}
The majority of the measured $d^2V/dI^2$ curves has a broad maximum whose position varies for different
contacts between 35 and 60\,meV, and whose FWHM is between 40 and 60 mV.
In general, the shape of these $d^2V/dI^2$  spectra is similar to the derivative of the resistivity
$d\rho/dT$, shown in Fig.\,\ref{fig:fig1} (left inset) and for that reason not shown here.
We would like to note that the differential resistance $dV/dI$ for these PCs increases by a factor
100--300\% from 0 to 100\,mV. This is typical for PCs in the thermal regime (see, e.g.
Refs.~\onlinecite{Naidyuk2005, Naidyuk2014}).
We have also calculated the $d^2V/dI^2$ curve expected in the thermal regime using Kulik's formula
(see Eq.\,(3.23) in Ref.\ \onlinecite{Naidyuk2005} and Refs. \onlinecite{Kulik1992,Verkin1993}).
The calculated curves have a shape similar to the $d\rho/dT$ shown in Fig.\,\ref{fig:fig1} (left inset).
The variation of the position of the maxima in $d^2V/dI^2$ for different PCs can be explained supposing
that the $\rho(T)$ in the PC core is modified due to an imperfect surface and to additional stress/perturbation
induced by the formation of the PC etc. We infer that in this case the current regime in the contact is thermal.

We suppose that, instead, the second type of spectra we observed correspond to a spectroscopic
(ballistic or diffusive) regime of the current flow in PCs.  Typical $d^2V/dI^2$ curves for such PCs
are shown in Fig.\,\ref{fig:fig2}. All these curves look similar to each other and display a clear
maximum at about 20\,mV with a subsequent background behavior. The relative change of $dV/dI(V)$
for such contacts is below 10 \%, i.e. one order of magnitude smaller than for the PCs in the thermal regime.
%{\bf \textcolor{blue}{Yura, please check. We have removed (commented) one short paragraph to avoid
%confusion of the reader and the referees with respect to supeconducting pecularities not considered in the present paper. }}
%{\bf LB: is this what you want to say?

\begin{figure}[t]
\includegraphics[width=0.45\textwidth]{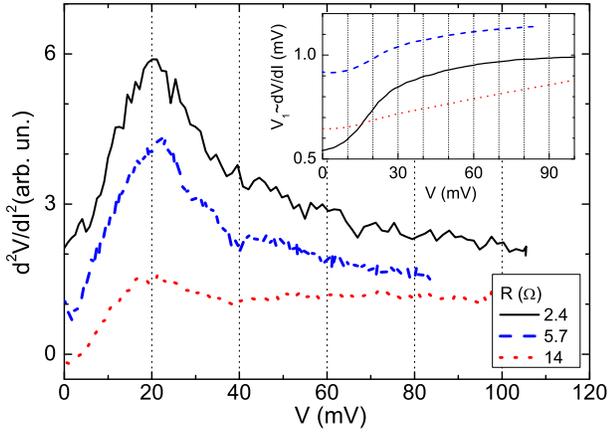}
\caption{ (Color online) PC spectra for three K122--Ag PCs with different resistance.
The curves are displaced vertically for clarity.  Inset: behavior of the differential resistance
for the corresponding curves in the main panel.
\label{fig:fig3}}
\end{figure}

In order to exclude that the 20\,meV feature we observed is due to extrinsic effects,
for instance caused by the Cu counterelectrodes, we have carried out measurements with both Cu and Ag needles.
The transverse phonons, which may in principle contribute to the PC spectra, are located between
15 and 20\,meV in Cu \cite{Naidyuk2005} and at 11--13\,meV in Ag.\cite{Naidyuk2005}
But, as clearly seen comparing Fig.\,\ref{fig:fig2} with Fig.\,\ref{fig:fig3}, in both sets of measurements
the maximum remains at 20\,meV, which indicates that this feature is solely due to intrinsic K122 excitations.

Usually phonon excitations are considered as the most likely candidates to explain peaks in PC spectra
in conventional superconductors. Their possible contribution in K122 will be considered
in the next subsection.

Spin fluctuations (or paramagnons) also give rise to characteristic features
in the PCS, as shown in Ref.~\onlinecite{Naidyuk1993}.
In K122, the inelastic neutron scattering spectra show a broad peak centered at $\omega_0 \sim 8$\,meV
at $T$ = 12~K \cite{Lee2011}. A very close estimate of the characteristic boson frequency
for spin fluctuations can also be extracted from the approximate scaling law: $T_c \approx 0.04T_{sf}$
for spin fluctuations (see Fig.\,1 in Supplement S1 of Ref.\ \onlinecite{Curro2005}).
This approximate relation is satisfied by many classes of {\it unconventional}, $d$-wave
like superconductors, such as heavy-fermion, Pu-based compounds, and high-$T_c$ cuprates.
In K122, with $T_c \approx$ 4\,K, expected $T_{sf}$ would be 100\,K ($\lesssim$ 10\,meV),
i.e.\ a factor two smaller than the position of the 20\,meV maximum in our spectra.
Therefore it is rather unlikely that the maximum in the PC spectra around 20\,meV is due to spin fluctuations.

Also tempting to relate our observations with recent STM/STS experiments \cite{Shan12,Wang2013},
which reported a strong dip in the tunneling $d^2I/dV^2$ spectrum at the energy 21.5${\pm}$0.8\,meV.
This feature was observed \cite{Wang2013} in two different systems, Ba$_{0.6}$K$_{0.4}$Fe$_2$As$_2$ and 
Na(Fe$_{0.975}$Co$_{0.025}$)As and it vanishes inside the vortex core or above $T_c$. 
This dip was attributed to the bosonic mode, which should appear in the tunneling spectra 
in the superconducting state of a strong coupling superconductor at an energy offset by the gap value.
In this case the bosonic mode energy would be about 14\,meV, that is not consistent with our 20\,mV maximum.
Interestingly, this dip is gradually smeared out with increasing $T$, but it does
not change the position.

ARPES  measurements have been reported in a recent study of
BaFe$_{1.9}$Pt$_{0.1}$As$_2$ (see Ref.\ \onlinecite{Ziemak2014}).
%In the analysis of the latter, two gap values, only,
%have been reported.
A bosonic energy of about 15\,meV can be estimated
from the peak-minimum distance from their data given in Fig.\,7(d).
This value is rather close to 14\,meV for the related systems mentioned above.
%\textcolor{blue}{ For a more complete collection of the PC work and discussion of PC measurements
%on other iron pnictides the reader is referred to the recent review paper by Gonnelli}
%\textcolor{blue}{ { \it et al.\ } \cite{Gonnelli2013}.}

\subsection{AB-INITIO CALCULATION OF THE EPI}

In order to estimate the contribution of phonons to the PC spectra we have calculated
the EPI properties of K122 {\em ab-initio} with density functional
perturbation theory, using plane-waves and pseudopotentials.~\cite{dfpt,qe,technical}.
The typical numerical accuracy of such calculations  amounts to $\sim 0.5$\,meV (6\,K)
for phonon frequencies, and to $\sim$ 10 $\%$ on the EP coupling constants.
We employed the structural data of Ref.~\onlinecite{Rosza81} -- space group $I4/mmm$.
%$a=3.842 \AA$, $c = 13.861 \AA$, $z_{As}$ = $0.3525$.
\begin{figure}[t]
\includegraphics[width=0.4\textwidth]{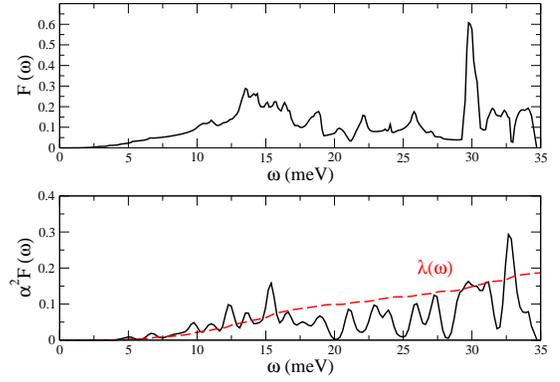}
\caption{ (Color online)
Upper panel: Phonon density of states $F(\omega$), calculated {\em ab-initio} using density
functional perturbation theory. Lower panel: Eliashberg function $\alpha^2F(\omega$)
(full line), and corresponding partial EP coupling constant $\lambda(\omega)$ (dashed line)-- Eq.\,(\ref{eq:lambda}).
The total $\lambda$ is 0.19, and the partial contribution from modes around 20\,meV is 0.05.}
\label{fig4}
\end{figure}
Fig.\,\ref{fig4} shows the calculated phonon density of states $F(\omega)$ and the EPI
spectral function $\alpha^2 F(\omega)$, defined as:
\begin{eqnarray}
\alpha ^{2}F(\omega ) &=&\frac{1}{N(0)N_{k}}\sum_{%
\mathbf{k,q},\nu }| g_{\mathbf{k,k+q}}^{\nu}| ^{2}\times
\notag \\
&&\delta (\varepsilon _{\mathbf{k}})\delta (\varepsilon _{\mathbf{%
k+q}})\delta (\omega -\omega _{\mathbf{q}}^{\nu }),
\label{eq:alpha}
\end{eqnarray}
where $N_{k}$ is the number of k-points used in the summation, $N(0)$ is the density of states
per spin at the Fermi level, and $\omega _{\mathbf{q}}^{\nu }$ are the phonon frequencies.
The EPI matrix element $g_{\mathbf{k}n,\mathbf{k+q}m}^{\nu}$ is defined by the variation
of the self-consistent crystal potential $V$ with respect to a frozen phonon displacement
according to the phonon eigenvector
 $e_{\mathbf{q}\nu }=\sum_{A\alpha }M_{A}\sqrt{2\omega_{\mathbf{q}\nu }}\epsilon _{A\alpha }^{\mathbf{q}\nu }u_{\mathbf{q}A\alpha }$.
The dashed line in Fig.\,4 (low panel) is the the value of the frequency-dependent
partially integrated EP coupling $\lambda(\omega)$:
\begin{equation}
\lambda(\omega)=2 \int_{0}^{\omega} d\omega \frac{\alpha^2 F(\omega)}{\omega}.
\label{eq:lambda}
\end{equation}

We first notice that the shape of the phonon density of states of K122
bares a strong resemblance to that of other 122 FeSC, which typically
extend up to $\sim$ 40\,meV. The high-lying modes are in-plane vibrations of Fe
and As atoms; out-of-plane vibrations are centered around 20\,meV, and the lowest
modes ($\lesssim 10$\,meV) have a substantial contribution from the K atom.
For the $A_{1g}$ and $B_{1g}$ modes at the $\Gamma$ point we obtain a frequency
of 24 and 27\,meV, respectively.
Our calculated spectrum has a Debye frequency of 261\,$\pm 6$~K, which is in quite
good agreement with the low-temperature specific heat data of Ref.~\onlinecite{Grinenko2014}
and~\onlinecite{Kittaka14, remdebye}, which report $\theta_D=276\pm 5$\,K
and $\theta_D$=274\,K, respectively. We expect that the theoretical Debye frequencies
would be even closer to experiment, if we took into account the slight lattice contraction at low temperatures.

The $\alpha^2 F(\omega)$, shown in the lower panel, exhibits a rich structure with
numerous peaks, as in the case of optimally doped pnictides and chalchogenides,
where the filling of the $d$ orbitals is near $d^6$. ~\cite{Boeri2008,Subedi2008}
Such a relatively unstructured spectral function, i.e. without a few especially pronounced maxima,
is characteristic of compounds with a  weak EP coupling. With respect to
those spectra, in K122 we  observe
a slight reduction of the coupling to the As out-of-plane modes centered
around 24\,meV, and a substantial increase of coupling to the
Fe and As in-plane modes at high frequencies ($\sim 30$\,meV).

The total  EP coupling constant, obtained by integrating Eq.\,(\ref{eq:lambda})
up to the highest frequency of the spectrum, is $\lambda_{tot}=0.19$,
i.e. comparable to the values obtained in the $d^6$ pnictides,~\cite{Boeri2008,Subedi2008}
and a factor three too low to explain the experimental $T_c$ of 3.5\,K.
The modes between 15 and 25\,meV contribute about one third of the total coupling constant,
i.e.\ $\lambda_{15-25}=0.05$. Including magnetic fluctuations would enhance this value
at most by a factor of two -- $\lambda_{15-25}=0.10$.~\cite{Boeri2010}
We believe that such an extremely low value would almost be invisible in the PC spectra,
also taking into account the kinematic factors which make $\alpha^2_{\rm PC} F(\omega)$
different from the ordinary  $\alpha^2F(\omega$) calculated here.
%~\cite{Gonnelli2010}
Notice also that in our PC spectra we do not resolve any of the other phonon modes
with smaller and larger energies, which in the calculations have a coupling comparable
or even larger than the out-of-plane modes. This is an indirect indication that
the effect of phonons on the PC spectra is indeed negligible. Our \textit{ab-initio}
calculations thus show that, at least from the point of view of standard
Eliashberg theory, the feature observed near 20\,meV cannot been ascribed to a phonon mode.
Thus we are forced to look for an alternative \textit{non-phonon} scenario, which we discuss below.

Before proceeding in our analysis, we would like to note that  a quantitative
theoretical model of EBI spectral functions in K122 should
also take into account the fact that for the $d$-wave superconductivity~\cite{Reid2012,Abdel2012,Grinenko2014}
the spectral functions responsible for the mass enhancement and
the pairing interaction \cite{Abdel2012} have different expressions.

\subsection{ALTERNATIVE NON-PHONON SCENARIO}

\begin{figure}[t]
\includegraphics[width=0.3\textwidth]{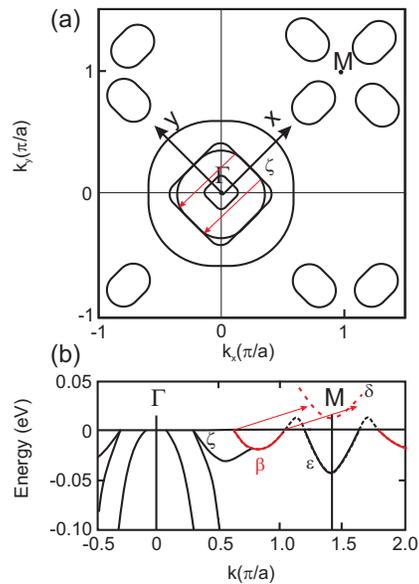}
.\hspace{1cm}
\caption{(Color online) Upper panel:
Schematic plot of the Fermi surface and band structure of K122. Red arrows: possible transitions (a)
back scattering of the electrons (b) creation of an indirect exciton. The effective band
structure sketched here has been adopted from recent ARPES measurements \cite{Sato2009,Yoshida2012} (see Fig.\,1 therein).
}\label{fig:fig5}
\end{figure}

We now suggest a possible alternative to phonons or spin fluctuations scenario.
We need to remind the reader about three specific features of the electronic structure
of the compound under consideration. First, K122 is nearly two dimensional.
Second, K122 is strongly hole doped -- i.e. Fe is in a nominal d$^{5.5}$ configuration,
so that one of the bands which are partially filled in optimally doped 122 systems and
form the electron pockets of the Fermi surface is unoccupied but still close to the Fermi-energy $E_F$.
 Third, one of the hole bands near the $\Gamma$-point has an almost square
 cross-section, as shown in  Fig.\,\ref{fig:fig5}.
For a thorough discussion of the electronic structure, including orbital-resolved
pictures of the Fermi surface, see e.g.\ Refs.~\onlinecite{OKA2011},~\onlinecite{haule} and \onlinecite{Valenti2014}.

% PLEASE LEAVE THIS OUT, IT WOULD TAKE HALF A PAGE TO  EXPLAIN THE FIGURE, I THINK IT'S BETTER WE USE DIMA'S  FIGURE!!!!

%\begin{figure}
%\includegraphics[width= 0.4\textwidth]{K_dima.eps}
%\vspace{}
%\caption{
%The shifted band structure in order to illustrate the similar curvatures of the involved bands and scattering moments.
%\textcolor{blue}{\bf Lilia, please, describe your figure in more detail, if necessary.}}|
%\label{shift}
%\end{figure}
%
Although there are quantitative differences between LDA and LDA+DMFT calculations~\cite{haule},
for example in the relative position and size of the hole pockets, and between these and
ARPES and de-Haas-van-Alphen experiments,~\cite{Sato2009,Yoshida2012,Terashima2013,Valenti2014},
these three basic features are robust.

In most FeSC around optimal doping, i.e.\ near $d^{6}$, the Fermi surface
comprises hole and electron sheets, centered around the $\Gamma$ and $M$ points of the Brillouin zone.
Two hole pockets derive from bands of $xz/yz$ character, which are degenerated at
the $\Gamma$ point, and the third has typically $xy$ or $3z^2 - r^2$ character.
Around the M point, two bands of $xy$ and $xz/yz$ character form two elliptical pockets,
with main axes along the ($k_x,k_y$) and ($-k_x,k_y$) directions.
In K122 ($d^{5.5}$), the Fermi level is shifted $\sim 0.2$\,eV below that of $d^6$.
The volume of the hole pockets is expanded, and that of the electron pockets is reduced.
At the M point, the bottom of the $xz/yz$ band, which forms the inner part of
the double-elliptical electron Fermi surface sheets for $d^6$, is raised a few tenths meV above the Fermi level.
Instead of a double ellipse, the Fermi surface now comprises 4 small ``propellers'',
along the ($k_x,k_y$)and ($-k_x,k_y$) directions.
At the same time, one of the $xz/yz$ hole pockets acquires a square-like cross-section.
A sketch of the electronic structure of K122, adapted from the ARPES data of Ref.~\cite{Sato2009,Yoshida2012},
is shown in Fig.\,\ref{fig:fig5}. The top panel (a) shows the Fermi surface,
with three large hole pockets around the $\Gamma$ point, and four ``propeller blades''
around the M point.  The two red arrows indicate the nesting vector of the square ($\zeta$)
Fermi surface ($q_{\zeta}=2 K_F^{\zeta}$).

The complicated electronic structure of K122 has important consequences on the PC spectra.
According to the theory of  PCS \cite{Kulik1977,Jansen1980},
the leading contribution to the energy dependence of the resistance in PCs
is the back-scattering of the quasiparticles due to their interaction with bosons. In this particular
case, the scattering from a point $-K_F^{\zeta}$ to a point $K_F^{\zeta}$ on the $\zeta$ Fermi surface sheet,
shown in Fig~\ref{fig:fig5}(a) is also compatible with the excitation of indirect excitons with $q_{\zeta}=2 K_F^{\zeta}$.

Indeed, Fig.\,\ref{fig:fig5}(b) shows that along the $x$ direction the shallow $xz/yz$ band
which forms the outer hole Fermi surface and the outer parts of the propellers can be almost
exactly translated on top of the unoccupied electron band $\varepsilon_{\delta}$,
for $q=2K^{\zeta}_F$ and $\epsilon_q \sim$ 20\,meV.
The portion of this shallow band which lies below $E_F$, denoted as $\varepsilon_{\beta}$
in the following, is strongly anisotropic, and concentrated in a small region of $\bf{k}$ and energy space.
As a consequence, the electron-hole excitations to the $\delta$ pocket are strongly peaked
in energy and momentum space.
This makes the possible charge excitation spectrum of K122 very different from that
of usual isotropic materials, where the scattering on the electron-hole continuum
leads to a featureless resistance contribution.

%{\bf LB: I wrote all scattering vectors (k,k') as vectors using
%mathbf; please check that this is correct.}
To find the contribution of these excitations to the PC spectrum we start from the
general expression for the back-flow current \cite{Jansen1980}:
\be
\Delta I \propto -2 \int_{0}^{\omega = eV} d\epsilon_1 \int_0^{\epsilon_1}
d\epsilon_2 \langle \langle \Gamma(\mathbf{k},\mathbf{k}') K(\mathbf{k},\mathbf{k}')\rangle_{\epsilon_1} \rangle_{\epsilon_2}
\label{eq:deltai}
\ee
$K(\mathbf{k},\mathbf{k}')$ is  a weighting factor depending on the incoming $\mathbf{k}'$
and outgoing momenta $\mathbf{k} = \mathbf{k}'+\mathbf{q}$;
note that $K$ is also affected by various properties of a real PC.
For the scattering of an electron inside the $\zeta$-band from one to the opposite side of
the Fermi surface (shown with red arrows in Fig.\,\ref{fig:fig5}(a)), for the sake of simplicity
we adopt the approximation $K(\mathbf{k},\mathbf{k}') =$ const., often found in literature.
The average in Eq.\,(\ref{eq:deltai}) is defined as:
$\langle ... \rangle_\epsilon = \sum_k \delta(\epsilon - \epsilon(\mathbf{k}))(...)$; here we assume $\hbar = e = 1 $.
Approximating the transition rate $\Gamma(\mathbf{k},\mathbf{k}')$
between an initial state $|k',i \rangle$ and a final state $|k,f \rangle$
with the probability of creating  indirect excitons with the energy
$\omega_\mb{p}\mb(q) =\epsilon_\delta(\mb{p-q})-\epsilon_\beta(\mb{p})$, we get:
%{\bf please check this formula; there are several indexes missing,
%and there is no $\mathbf{p}$ wave-vector; what is $\nu$?}
\be
\frac{d^2 V}{dI^2} \propto \sum_\mb{p} \sum_{\mathbf{k, q}}\delta(\omega
- \eps_\mb{k}) \delta(\eps_\mb{k+q})  \delta(\epsilon_\zeta(\mb{k}) -
\epsilon_\zeta(\mb{k+q}) - \omega_{\mb{p}}(\mb{q}) )
\label{eq:dr}
\ee
We now introduce analytical approximations for the dispersion of the relevant bands.
In particular, for the $\zeta$ band we assume a linear dispersion:
 $\epsilon_\zeta(\mb{k}) = v_{F_0} (k_x-K^{\zeta}_{F})$;
for the $\delta$ band a parabolic spectrum, centered around the $M$ point:
 $\epsilon_\delta(\mb{k} + \mb{K}_M) = k^2/2m - \mu + E_a$.
Furthermore, we approximate the dispersion of the $\beta$ electron-like pocket, centered around
 $\bar{\mb{Q}} = \mb{K}_M - 2 \mb{K}_F^{\zeta} + \Delta q$,
and extending in the $k_x,k_y$ direction,
as $\epsilon_{\beta}(\mb{k}+\bar{\mb{Q}}) = k^2_x/2m - k^2_y/2m^* -\mu$.
We further assume $m^* \gg m$ and neglect the $k^2_y/2m^*$ contribution to the kinetic energy.
A straightforward calculation yields:
\begin{eqnarray}
\frac{d^2V}{dI^2} &\propto&  \frac{1}{\omega  +  v_{F_0}/v_{F} \Delta E} \nonumber \\
 &\times&   Re \left[
\left[ \omega \left( 1 + \frac{v_{F}}{v_{F_0}}\right) -E_a +\Delta E \right. \right.
\nn \\ && \left.\left. \,\,\,\,\,\,\,\,\,\,\,\,\,\,\,\,\,\,\,\,\,\,\,\,\,\,\,\,\,\,\,\,\,\,
+\frac{1}{4E_0}
%\alpha
\left(\omega + \frac{v_{F_0}}{v_{F}}\Delta E\right)^2\right]^{1/2} \right. \\
&& \left. - \left[\omega\left(
 1 - \frac{v_F}{v_{F_0}}\right) -E_a-\Delta E\right. \right. \nn \\
 && \left.\left. \,\,\,\,\,\,\,\,\,\,\,\,\,\,\,\,\,\,\,\,\,\,\,\,\,\,\,\,\,\,\,\,\,\,
 +
 %\alpha
 \frac{1}{4E_0}\left(\omega + \frac{v_{F_0}}{v_{F}}\Delta E\right)^2\right]
\right] \bigg] \textcolor{blue}{,}
\nn
\label{eq:deltaex}
\end{eqnarray}
where $v_F = p_F/m$ is the Fermi-velocity in the band $\beta$, $\Delta E = v_F \Delta q$
and $E_0 = 0.5 mv_{F_0}^2$.
The charge excitations (excitons) give rise to a differential
conductivity which starts abruptly at $\omega \sim E_a$, and decays at high energies as $1/\sqrt{\omega}$.
This expression gives a very good fit of the behavior of experimental PC spectra at large frequencies.

In order to fit the experimental spectra also at low frequencies,
we consider an additional contribution to the back-flow, due
to spin fluctuations;  the empirical spectral density is given by
\begin{equation}
B(\omega) = \frac{\omega_0}{\pi} \frac{\omega}{\omega^2 +\omega^2_0} \quad.
\label{eq:SFB}
\end{equation}
Here, we set $\omega_0 = 8$\,meV, from the inelastic neutron
scattering data at $T$ = 12~K \cite{Lee2011,Wang2013}.
For the present analysis, we are more interested in the asymptotic behavior of $B(\omega) \propto \omega^{-1 } $ for $\omega \gg \omega_0$,
which translates into a $1/\omega$ decay of the differential conductivity than in the the precise value of $\omega_0$.
Thus, the total differential conductivity reads:
\begin{eqnarray}
\frac{d^2V}{dI^2} &=&
A \left[B \frac{\omega_0}{\pi}\frac{\omega}{\omega^2+\omega_0^2}+ \right.
\label{eq:fit}
\nonumber
\frac{1}{\omega+ v_{F_0}/v_{F}\Delta E} \\ &\times & Re \left[
\left[ \omega \left( 1 + \frac{v_{F}}{v_{F_0}}\right) -E_a +\Delta E \right.
\right. \nn \\ && \left.\left. \,\,\,\,\,\,\,\,\,\,\,\,\,\,\,\,\,\,\,\,\,\,\,\,\,\,\,\,\,\,\,\,\,\,
 +\frac{1}{4E_0}\left(\omega + \frac{v_{F_0}}{v_{F}}\Delta E\right)^2\right]^{1/2} \right. \\
&& \left. - \left[\omega\left(
 1 - \frac{v_F}{v_{F_0}}\right) -E_a-\Delta E\right. \right. \nn \\
 && \left.\left. \,\,\,\,\,\,\,\,\,\,\,\,\,\,\,\,\,\,\,\,\,\,\,\,\,\,\,\,\,\,\,\,\,\,
 +\frac{1}{4E_0}\left(\omega + \frac{v_{F_0}}{v_{F}}\Delta E\right)^2\right]
\right] \bigg] \textcolor{blue}{,}
\nonumber
\end{eqnarray}
where $A$, $B$ are multiplicative factors which account respectively
for the (unknown) area of the PC and for the relative weight of
the excitonic and spin-fluctuation contribution to the spectral function;
$E_0=$~120\,meV.

\begin{figure}[t]
\includegraphics[width= 0.45\textwidth]{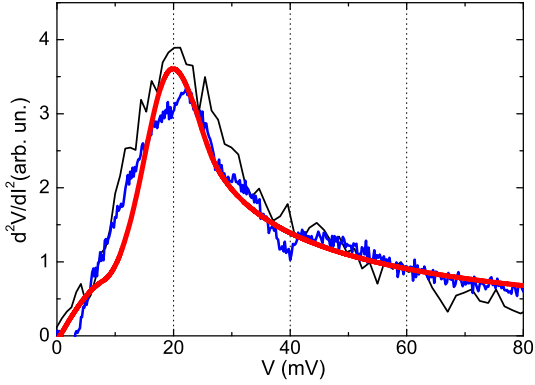}
\caption{ (Color online)
PC spectra from Fig.\,\ref{fig:fig3} (black and blue).
A resonable description of these spectra is obtained by Eq.\,(\ref{eq:fit}) (red)
adopting the following parameter values:
%\textcolor{blue}{obtained by this procedure} %of the fit are:
$v_F/v_{F_0}=0.3$ and $B= 0.25$, $E_a  = 18$~meV, $\Delta E  = - 3$~meV,
%$\alpha=0.03$meV$^{-1}$
$E_0=$~120~meV,
and  $A$=22.}
%PC spectra from Fig.\,\ref{fig:fig3},
%fitted with Eq.\,(\ref{eq:fit}), with (upper panel, dashed line),
%and without (lower panel, dotted line) a spin-fluctuation term.
%The parameters of the fit are: Upper panel:
%$v_F/v_{F_0}=0.93$ and $B= 1$, $E_a  = 18$\,meV, $\Delta E  = - 11$\,meV,
%and  $A$=12, 8.6, 3.2 arb. un., respectively,
%for the black, blue and red curves. Lower panel: $B= 0$, $E_a  = 20$\,meV, $\Delta E  = - 14$\,meV,
%and  $A$=14, 10, 3.8 arb. un. (black, blue, red curves).}
\label{fig:fig6}
\end{figure}

Representative examples of the measured second derivatives of the $I-V$ curves
are shown in  Fig.\,\ref{fig:fig6} together with fits through Eq.\,(\ref{eq:fit}).
The experimental data correspond to two K122--Ag spectra with different resistance,
the fitting parameters are reported in the caption of Fig.\,\ref{fig:fig6}.

In all cases, the experimental spectra show a clear 1/$\sqrt{\omega}$
decay at high energies, which is not captured by the spin-fluctuation
term; this behavior is also incompatible with phonons, which
have a finite spectrum, and thus PC spectrum has constant value \cite{Naidyuk2005} above Debye energy
%a cut-off in the differential conductivity
of $\sim$ 40\,meV.
On the other hand, the excitonic contribution Eq.\,(\ref{eq:deltaex})
has a 1/$\sqrt{\omega}$ tail.
A finite spin-fluctuation term ($B \neq 0$) in Eq.\ref{eq:fit} is needed
to account for the spectral weight at low frequencies ($\omega \leq E_a$).

%The deflection above 40\,meV in spectrum of
%PC with $R=14\,\Omega$ (lower curves) might be attributed to
%\textcolor{red}{deviation}\textcolor{blue}{s} \textcolor{red}{from}
% the ballistic regime considered here.
%{\bf \textcolor{blue}{Yura, the decision to remove the third curve or not is left
%for you. In our opinion the deviation  above 40 meV is well enough explained.
%LB: I don't like too much the third curve (large R); the fit is not really good!}}

%The lower panel of Fig.\,\ref{fig:fig6} shows the same PC spectra fitted with only the
%exciton contribution (Eq.\,(\ref{eq:deltaex})); the formula captures quite well
%the behavior of the experimental data at large-$\omega$'s, but the spectral weight at low
%frequencies is missing. This can only be recovered including a
%spin fluctuation term ($B \neq 0$) in Eq.\,(\ref{eq:fit}).

The charge excitation (excitonic) mechanism proposed here
has clear fingerprints, distinct from usual phonon excitations, which should be
easy to detect experimentally.
We propose a few experimental tests, which would definitely confirm
our scenario and rule out the possibility that the
20\,meV peak is due to coupling to $c$-axis phonons.

First of all, our model predicts a strong dependence of the peak position
on $E_a$, which is shifted to higher(lower) values upon hole(electron) doping.
This is in stark contrast to what is expected for phonons, since it is known
that the phonon frequencies in pnictides depend only weakly on the doping.
%{\bf LB: I inserted the follwing sentence}
PC studies of samples with different dopings could then be used to check our scenario.

Another possibility is to use high resolution EELS (in reflection)
\cite{Ibach1993} probing the longitudinal density-density response
i.e.\ the in-plane polarizations: this would help to distinguish
%in this case, the
%(in the case of nonmagnetic singlet excitons, for magnetic triplet excitons magnetic INS
%scattering should be considered instead (see below)),
$c$-axis polarized phonon modes at $\sim 20$\,meV from the in-plane
low-lying non-vertical interband transition proposed here. Its high resolution, up to
0.5\,meV, may also be able to reveal further details of the electronic structure.
Note that EELS would be able to detect nonmagnetic singlet excitons,
while to detect magnetic triplet excitons spin-polarized
inelastic neutron scattering should be used instead.

More theoretical and experimental studies are required to clarify the interplay of
charge excitations with other degrees of freedom in (doped) K122.
For example, in doped samples, it would also be essential to understand the effects
of deviations from stoichiometry, on disorder, and/or electronic correlations on
the PC spectra; this is particularly crucial in FeSC due to the contiguity of an orbital
selective Mott transition.~\cite{Mottsel,Medici2012}
These measurements may also help to gain further insight on the strength of the EPI
and details of the total mass renormalization and their interplay with $d$-wave
superconductivity.

This interplay depends on the magnetic nature of the excitons under consideration.
In case of non-magnetic singlet excitons, if $d$-wave superconductivity is suppressed due to
a strong enough pair-breaking disorder, the residual superconductivity, if any at all, might
be of \ the $s_\pm$-wave type.
In this case, the excitonic mechanism proposed here would act as a non-flipping-spin
{\it charge} excitation process. Like phonons, it would therefore compete with spin
fluctuations in case of $d$-wave superconductivity, and \textit{support} superconductivity
in case of an $s_\pm$ pairing regime with accidental nodes, or
another complex pairing-regime induced by disorder.

%{\bf I introduced a new sentence, LB}
We will now try to estimate the contribution of excitons to the total coupling
constant for the EBI.

Since in clean samples a transition to an $s_\pm$ pairing regime has not been observed,
nonmagnetic excitons compete with spin-fluctuations and therefore we expect:
\begin{equation}
\lambda_{\rm ph} +\lambda_{\rm exc} < \lambda_{\rm sf} .
\label{eq:lambdaex}
\end{equation}

Indeed, several experimental evidences point to a $d$-wave symmetry in clean samples.
For example, the low-$T_c$  of K122 fits very well the empirical relation for
several $d$-wave superconductors \cite{Moriya2003,Curro2005}:
\begin{equation}
T_c\approx 0.04T_{\rm sf} , \
\label{eq:tc}
\end{equation}
if we use for $k_{\rm B}T_{\rm sf}=\hbar \omega_0=8$\,meV $\approx 93$\,K,
extracted from inelastic neutron scattering measurements ~\cite{Lee2011} and $T_c=$~3.6~K.
~\cite{Grinenko2014}

%%% SORRY, this sentence is too long and it is not clear to me how
%%% IT IS RELATED TO THE PREVIOUS DISCUSSION
%The value of $T_{\rm sf}$ used here was
%estimated from the position of the peak of the
%Im$\chi (\omega, \mathbf{Q})$, measured by inelastic
%neutron scattering, at $\mathbf{Q}=\mathbf{\bar{Q}}$.~\cite{Lee2011}
%Here,
%$\mathbf{\bar{Q}}$ is an incommensurate "propagation" vector
%which points to the existence of a still unknown magnetic phase
%different from the commensurate phase of the parent compounds
%Ba-122 and Sr-122.
%
%The critical temperature $T_{c,0}\approx 3.7$~K from
%Eq.~\ref{eq:tc} is very
%close to the extrapolated clean limit
%value of 3.6~K recently estimated for K122 \cite{Grinenko2014}.

In case of  $s_{\pm}$ wave symmetry with accidental nodes on one sheet of the Fermi surfaces,
only, one would expect a significantly higher  $T_c$, intermediate between the $d$-wave
value and that of a nodeless $s_{\pm}$ gap.
%~\cite{Drechsler2013}

Substituting our calculated $\lambda_{\rm ph} \approx 0.15-0.2$, and $\lambda_{\rm sf} \approx 0.6$ to 0.8
 deduced from a previous Eliashberg analysis \cite{Abdel2013,Grinenko2014} into Eq.\,(\ref{eq:lambdaex})
we can estimate an upper bound for $\lambda_{\rm exc} < 0.4-0.45$.
In  this estimate it is assumed that we deal with singlet-excitons, only.

In the case of triplet excitons the observed peak would be a special type of spin fluctuation
visible in the magnetic inelastic neutron scattering spectra at low temperature.
%%% LB: Sorry, I removed this sentence because to me it looks in contraddiction
%%% with the next (wang measurements)
%For the moment, the only available experimental data  are restricted to energies below 15~meV
%and $T=12$~K, only.~\cite{Lee2011} Additional low-\textit{T} INS-measurements around 20 meV
%at $T$ slightly above $T_c$ would be highly desirable.
In the triplet case the magnetic excitons would support a $d$-wave type of superconductivity.
In this context very recent inelastic neutron scattering data obtained by Wang {\it et al.}
\cite{Wang2013b} at $T=5$\,K for K122 are of interest.
According to these authors there is practically no spectral weight above about 20\,meV
for magnetic excitations. Then one might conclude that a magnetic exciton scenario is
rather unlikely.
%(Strictly speaking studies of the doping, pressure, and disorder effects on $\lambda_{\rm exc}$
%and the excitonic feature in general, provide an interesting and challenging issue
%for future experimental and more sophisticated theoretical work. In the case of further hole
%doping, clear upshifts in its energy are expected, but the corresponding change of the coupling constant
%$\lambda_{\rm exc}$ remains elusive at present.

\section{CONCLUSION}

We have investigated PC spectra in the normal state of the low-$T_c$ iron-pnictide compound
K122. A single maximum at about 20\,meV has been observed.
%Taking into account the existing Raman, ARPES and STM/STS data on similar 122 compounds, it
%might be tempting to connect the maximum in the PC spectra with the $A_{1g}$ and $
%B_{1g}$ phonon modes, derived from the out-of-plane vibrations of As and Fe atoms.
%These phonons can induce antiferro-orbital fluctuations, which would however give rise to
%an $s_\pm$-wave superconducting state \cite{Kontani2012} or to an $s_\pm$-wave state
%with accidental nodes \cite{Okazaki2012}, both of which have been not observed,
%so far \cite{Reid2012,Abdel2013}.
In this work, we have proposed a novel %alternative
\textit{nonphonon} and \textit{nonmagnetic }scenario to interpret the PC spectra,
based on the presence of unoccupied electron bands close to Fermi energy.

This scenario is strongly supported by a DFT linear response calculation of the Eliashberg EPI
function and of the corresponding EP coupling constant $\lambda$, which shows an
extremely low coupling for phonon modes between 15 and 25\,meV, comparable or lower
to that of the remaining phonon spectrum ($\lambda_{15-25}$=0.05, $\lambda_{tot}=0.19$).

Our work provides the first evidence for an additional bosonic excitation in FeSC,
beyond the ones usually discussed in literature - phonons, spin and orbital fluctuations.
To the best of our knowledge, this is the first time that this type of excitonic
charge excitation (indirect longitudinal excitons) is reported for a metallic system.
In typical metals the difficulty to observe excitons is usually ascribed to the large dielectric
screening provided by the fast conduction electrons. However, in the present
somewhat "anomalous" case, where heavy charge carriers are present with
large mass renormalizations,
this detrimental screening might be significantly suppressed.

The present finding extends the list of exceptional cases in which PCS could detect
the interaction of electrons with bosons other than phonons.
These include magnons \cite{Akim}, crystal field excitations interacting with conduction
electrons in the magnetic superconductor \cite{Naidyuk2007} HoNi$_2$B$_2$C, in the superconducting
heavy fermion system PrOs$_4$Sb$_{12}$ \cite{Kvitnitskaya2006},  in  PrNi$_5$ \cite{Reiffers1989},
and with paramagnons in the nearly ferromagnetic  CeNi$_5$.\cite{Naidyuk1993}

\section*{Acknowledgements}
Yu.G.\ N.\ and O.E.\ K.\ thank the IFW Dresden for hospitality and the
Alexander von Humboldt Foundation for financial support. Funding by
the National Academy of Sciences of Ukraine under project $\Phi $3-19
is gratefully acknowledged. D.E., S.W., L.B. and S.-L.D. acknowledge the Deutsche
Forschungsgemeinschaft DFG (priority program SPP 1485) for support. S.W.\ thanks also funding by
the research training group GRK 1621 as well as by the project WU 595/3--1.
Discussions with S.V. Borisenko, R.S.\ Gonnelli, S.\ Johnston, M.\ Knupfer, Yu.A.\ Kolesnichenko,
A.N.\ Omelyanchouk and H.\ Rosner are kindly appreciated. The assistance of N.L.\ Bobrov
and K.\ Nenkov during some experiments is acknowledged.


\begin{thebibliography}{99}
\bibitem{footnote2}
Thus, in general, the PC EPI function is quantitatively different from the usual
Eliashberg EPI function responsible for  the superconductivity (see Eq.\,(\ref{eq:alpha})).
However, both functions are strongly enhanced at frequencies corresponding
to bosons with a strong interaction to electrons.
\bibitem{technical}
All calculations were performed in the generalized gradient approximation
\cite{DFT:PBE} using plane-waves and ultra-soft pseudopotentials~\cite{qe,Vanderbilt}.
We employed a cut-off of 40 (480) Ryd for the wave-functions (charge densities).
The electronic integration was performed over an $8^{3}$ $\mathbf{k}$-mesh with a $0.01$ Ry
Hermitian-Gaussian smearing. Finer grids ($20^{3}$) $\mathbf{k}$-points
were used for Fermi surface averages of the EP matrix elements in \ref{eq:alpha}.
Dynamical matrices and EP linewidths were calculated on a $4^{3}$ uniform
grids in $\mathbf{q}$-space, and Fourier-interpolated to obtain dispersions in the whole Brillouin zone
\end{thebibliography}
\end{document}